# Contrasting Effects of Functionalization in Binary and Medium-Entropy MXene Coatings for Corrosion Protection

Aqsa Fayyaz*, Ulises Martin Diaz[#], Jianyu Dai*, Homero Castaneda[#†], Chenglin Wu[†*]

*Department of Civil and Environmental Engineering, Texas A&M University, College Station, TX, 77843, USA.

[#]Department of Materials Science and Engineering, Texas A&M University, College Station, TX, ,77843, USA.

[†]Corresponding authors: hcastaneda@tamu.edu; chenglinwu@tamu.edu

## Abstract

Developing scalable and environmentally benign anticorrosion coatings is critical for protecting steel infrastructure in chloride-rich environments. Here, a nacre-inspired multilayer epoxy coating reinforced with four MXene systems is investigated. This architecture forms a dense lamellar network that increases diffusion tortuosity and introduces electroactive surfaces for ion interactions. Electrochemical impedance spectroscopy (EIS) confirms that the multilayer design increases coating resistance from ~$10^3$ to ~$10^8$ $\Omega \cdot cm^2$. A clear performance hierarchy was observed: $P$-$(TiVCrMo)C_3$ > $O$-$Ti_3C_2T_x$ > $O$-$(TiVCrMo)C_3$ > $P$-$Ti_3C_2T_x$. Density functional theory (DFT) calculations reveal that $P$-$Ti_3C_2$ strongly adsorbs $O_2$, indicating higher surface reactivity, while oxygen termination stabilizes the surface by partially passivating Ti sites. In contrast, $P$-$(TiVCrMo)C_3$ exhibits strong adsorption of oxygen-containing species due to its multi-metal electronic structure, promoting the formation of protective oxides. These results highlight the delicate balance of surface chemistry, electronic structure, and compositional entropy in designing next-generation MXene-based anticorrosion coatings for marine and industrial environments.

**Keywords:** $Ti_3C_2T_x$; $(TiVCrMo)C_3$; carbides; corrosion protection; Medium entropy MXene; density functional theory.

# Graphics for Manuscript

**1. Bare steel: Localized Pitting Corrosion**

**2. Coated Steel: Suppressed corrosion**

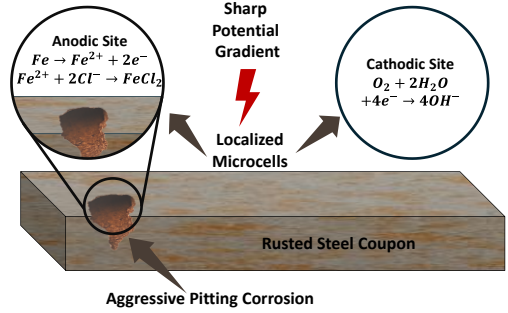
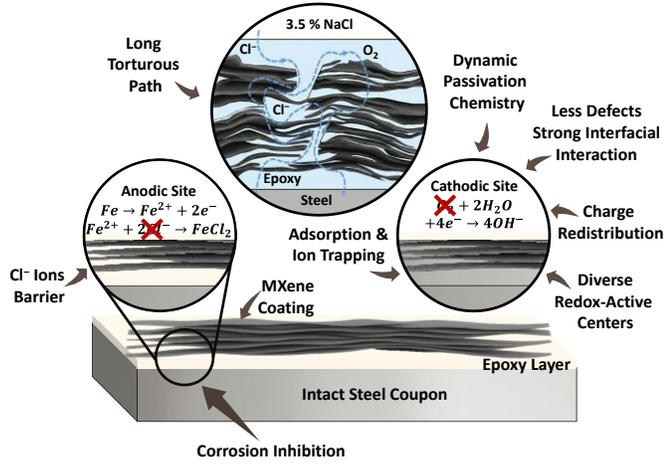

# 1 Introduction

Corrosion remains one of the biggest curses of modern infrastructure and structural metals, imposing an estimated annual cost of $2.5 trillion globally and $276 billion in the United States alone [1]. Most metal structures corrode because of formation of porous iron oxide (rust) layer which dissolves in water and facilitates the ingress of water, and oxygen. Moreover, in saline environment chloride ions, further penetrates and destroys the oxide layer, accelerating degradation [2]. Therefore, corrosion resistant methods must be employed to counteract this degradation. Currently, cathodic protection [3], corrosion inhibitors [4] and organic coating [5] are primarily used. Among these organic coating has been extensively explored due to ease of application, cost effectiveness, robust mechanical integrity and good adherence to variety of metal substrates [6]. However, during the curing process microdefects and cracks inevitably appear due to cross-linking and evaporation of solvents, providing initiation points for the ingress of aggressive ions to the metal substrate [7]. Hence, necessitating the use of nanofillers to reinforce the matrix and enhance corrosion resistance.

MXene are an emerging class of two-dimensional (2-D) materials and are typically synthesized by selectively etching the A element from the respective MAX phases (M is an early transition metal, A is IIA or IVA element and X is either carbon or nitrogen making it carbide or nitride) [8, 9]. $Ti_3C_2T_x$ is one of the most studied MXene and it consists of three layers of Ti atoms and two layers of C atoms alternating with each other with $T_x$ representing the surface terminal group such as –OH, –O and –F and their percentage depend upon the method of exfoliation deployed [10]. Owing to its exceptional metallic conductivity of 24,000 S cm$^{-1}$, high surface area, a young's modulus of around 334 GPa, tunable surface chemistry makes it the suitable candidate for multifunctional applications [11].

Comprehensive studies over the last decade have explored the use of MXene based coating across diverse systems. Song et al. [12] fabricated a $Ti_3C_2T_x$-phytic acid-gallic acid (M–P–G) hybrid composite that achieved an impedance modulus of $1.51 \times 10^9$ Ω $cm^2$, surpassing conventional coatings by two orders of magnitude, owing to the coordinated protection of gallic and phytic acids. Similarly, Li et al. [13] developed M16@MXene–$SiO_2$ epoxy hybrid coatings, achieving enhanced barrier protection with impedance values up to $5.98 \times 10^9$ Ω $cm^2$. In another study, Ning et al. [14] employed Ce(III)-modified $Ti_3C_2T_x$ in epoxy, which improved corrosion resistance by promoting compact passive-film formation ($|Z| \approx 7.2 \times 10^9$ Ω $cm^2$). Likewise, Ji et al. [15] prepared APTES-modified MXene@GO QDs epoxy coatings, achieving impedance of $4.4 \times 10^9$ Ω $cm^2$ and superior adhesion.

Functionalization of MXene with such components has effectively improved charge-transfer resistance and extended corrosion protection. Nevertheless, most reported approaches rely on complex and multi-step modification processes, involving diverse surface chemistries that limit scalability and economic feasibility. In addition, research to date has mainly centered on binary or ternary MXene with the influence of multicomponent MXene chemistries remaining largely unexplored. Inspired by the hierarchical architecture of Nacre, consisting of 95 vol% of stiff aragonite platelets embedded in soft organic matrices a multilayer composite coating was developed [16]. The design features alternating epoxy and nanomaterial layers, where the epoxy serves as a flexible matrix and the nanomaterials act as reinforcing barriers to extend diffusion paths and suppress crack propagation. Since the corrosion resistance of MXene-based coatings strongly depends on surface chemistry and structural order, this study systematically compares four multilayer nanomaterials coatings i.e., *P*-$Ti_3C_2T_x$, *O*-$Ti_3C_2T_x$, *P*-$(TiVCrMo)C_3$, *O*-

(TiVCrMo)C$_3$ to evaluate the effect of oxidation and compositional complexity [17, 18]. Density functional theory calculations were also performed to further support these observations.

## 2 Materials and Methods

### 2.1 Preparation of coating samples

C1010 carbon steel coupon was selected as a corrosion substrate with an exposed area of 1 inch × 1 inch and remaining part was masked with the tape. The surface of the coupon was polished with sandpaper from 40 to 1200 grit size, sonicated in DI water and ethanol separately for 30 min each to remove the residual polishing material and subsequently dried in oven for few hours.

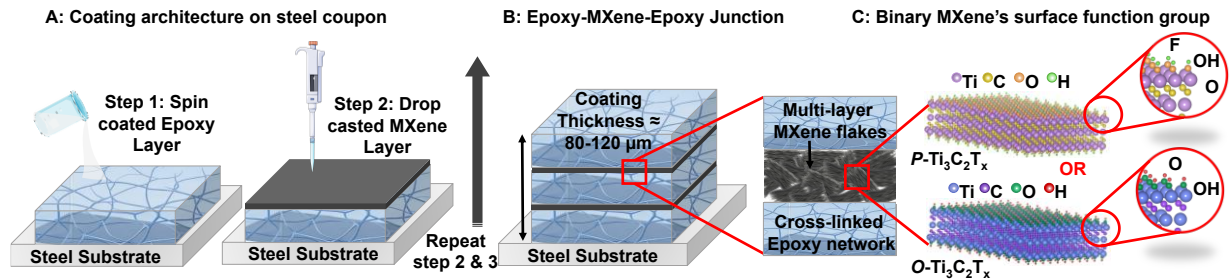

*Figure 1: Schematic illustration of the fabrication process of multilayer coating, illustrating hierarchical organization from macroscopic layer assembly to microscopic interfacial interactions and atomic-scale surface functionalization.*

To prepare the coating, the suspension of all the four MXene *P*-(TiVCrMo)C$_3$, *O*-(TiVCrMo)C$_3$, *P*-Ti$_3$C$_2$T$_x$, *O*-Ti$_3$C$_2$T$_x$ was prepared at a concentration of 5 mg/ml in DI water. Each MXene was dispersed separately in Eppendorf vials and sonicated for 30 min. Bare steel coupons were used as substrates for constructing single and multi-layer coatings, consisting of epoxy as the outer layers and MXene as interlayers as illustrated in schematics (c) of Error! Reference source not found.. The first epoxy layer was deposited by spin coating at 2500 rpm for 60 s, followed by curing time of at least 12 h. Previously sonicated MXene suspension was drop-casted on cured epoxy surface and cured for12–15 h before applying the next epoxy layer as demonstrated in **Figure 1**. This deposition, curing cycle was repeated to fabricate three types of samples: single layer (1L; only epoxy on bare steel), three-layer (3L; epoxy/MXene/epoxy) and five-layer (5L; epoxy/MXene/epoxy/MXene/epoxy) coatings.

## 3 Results and Discussion

### 3.1 Characterization

The cross-sectional SEM image in **Figure 2(a)** reveals a stratified epoxy/$P$-Ti$_3$C$_2$T$_x$ architecture deposited on a steel substrate, where three epoxy layers alternate with two Ti$_3$C$_2$T$_x$ layers. The clear contrast between the epoxy and MXene regions confirms the successful construction of a periodic multilayer structure. EDS mapping further substantiates this arrangement: the Fe signal in **Figure 2(b)** is confined to the steel substrate, while the Ti signal **(Figure 2(d))** distinctly delineates the MXene layers embedded within the epoxy matrix. A top-view SEM image in **Figure 2(f)** of the $P$-Ti$_3$C$_2$T$_x$ surface highlights the characteristic flake-like morphology and high-magnification SEM image in **Figure 2(g)** of an individual flake (2 µm) further confirms the accordion-like morphology, providing direct evidence of successful MXene exfoliation. This hierarchical configuration, integrating ordered layering with nanosheet morphology, is expected to play a critical role in enhancing barrier performance and corrosion resistance [19].

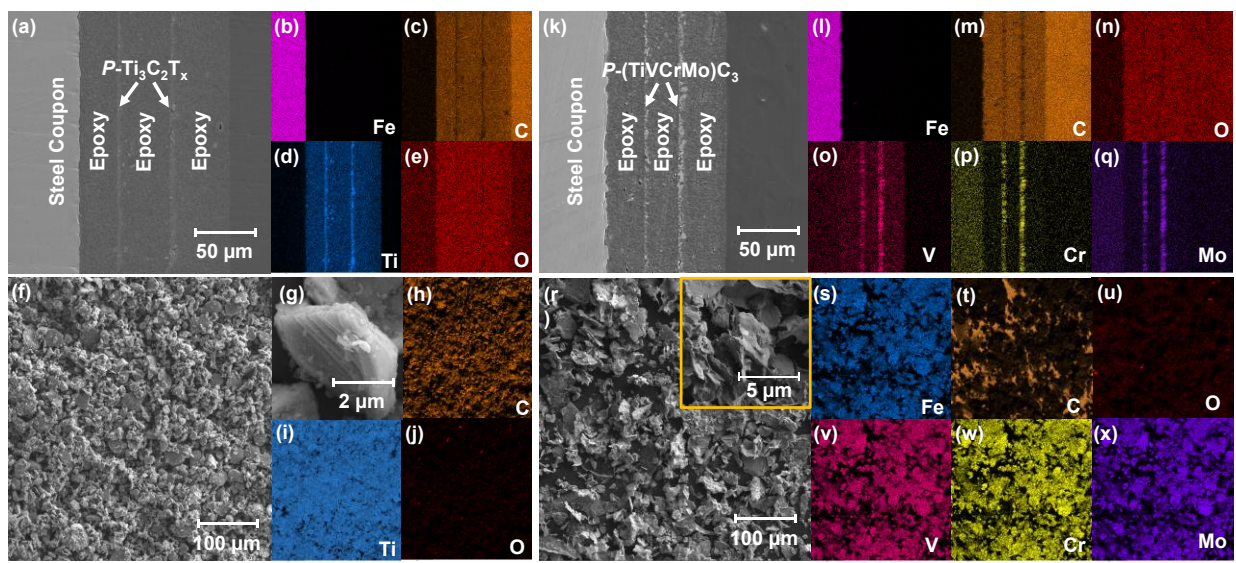

*Figure 2: SEM micrographs and corresponding EDS analyses of fabricated multilayer coatings: (a) cross-sectional view showing the five-layer architecture of P-Ti$_3$C$_2$T$_x$; (b–e) corresponding EDS elemental maps; (f) top-view morphology of the P-Ti$_3$C$_2$T$_x$ layer with (g–j) elemental distribution; (k) cross-sectional SEM image of the P-(TiVCrMo)C$_3$ coating and (l–q) corresponding EDS maps; (r) top-view morphology of the P-(TiVCrMo)C$_3$ layer with (s–x) EDS mapping.*

Analogously, the cross-sectional SEM image of the *P*-(TiVCrMo)C$_3$ in **Figure 2(k)** coating reveals a stratified multilayer configuration. The compositional distinction is corroborated by EDS mapping, where V, Cr, Mo, signals delineate the medium entropy MXene-rich layers. The top-view SEM image **Figure 2(r)** of *P*-(TiVCrMo)C$_3$ shows flakes with a comparatively rougher and more irregular morphology than *P*-Ti$_3$C$_2$T$_x$, with thicker edges. Inset of **Figure 2(r)** displays the individual flakes of *P*-(TiVCrMo)C$_3$ and exhibits pronounced crumpling and wrinkling, a morphology that reflects the intrinsic structural complexity of medium-entropy MXene, where multielement composition induces enhanced surface corrugation and sheet rigidity [20]. A similar cross-sectional and surface morphology for the *O*-Ti$_3$C$_2$T$_x$, and *O*-(TiVCrMo)C$_3$ coatings are presented in **Figure S2(a-j)** and **(k-x)** respectively.

AFM analysis was performed to examine the surface morphology and thickness of the four MXene–epoxy coatings: *P*-Ti$_3$C$_2$T$_x$, *O*-Ti$_3$C$_2$T$_x$, *P*-(TiVCrMo)C$_3$ and *O*-(TiVCrMo)C$_3$. **Figure 3(a-d), S3 (a-d)** presents the corresponding topography images and height profiles. All samples exhibit a distinct step between the epoxy layer and the deposited MXene film, confirming the successful formation of uniform and continuous coatings. The measured step heights indicate film thicknesses in the range of 2.2–2.6 µm, indicating the smooth interface, homogeneous coverage and uniform dispersion of MXene over the waterborne epoxy surface.

X-ray diffraction (XRD) patterns of Ti$_3$AlC$_2$, *P*-Ti$_3$C$_2$T$_x$, and *O*-Ti$_3$C$_2$T$_x$ are shown in **Figure 3(e)**. The Ti$_3$AlC$_2$ precursor exhibits characteristic MAX reflections at 2θ ≈ 9.6° (002), 19.2° (004), 34° (101/103), and a strong peak at 39–41° corresponding to (104)/(105)/(106). After selective Al removal, the *P*-Ti$_3$C$_2$T$_x$ displays the typical MXene (00l) reflections, with the (002) peak shifted to ~8.3° and a weaker (006) reflection near 18°, while the absence of the MAX (104) at ~39–41° confirms successful etching. Upon oxygenation, the (002) peak shifts slightly further to lower angle (~8.15°), suggesting an enlarged interlayer spacing due to oxygen terminations and intercalated species, while higher-order (004) and (006) reflections emerge at ~16.4° and ~24.6°, respectively, consistent with MXene stacking. Additional peaks appear at 34°, 36.7°, 41.0°, and 60.2°, which can be indexed to TiO$_2$ impurities, confirming partial oxidation of the MXene surface.

**Figure 3(f)** exhibits the XRD pattern of (TiVCrMoAl)C$_3$, *P*-(TiVCrMo)C$_3$ and *O*-(TiVCrMo)C$_3$. The sharp reflections at ~15°, 37°, 39°, and 40–42°, corresponding to the (002), (104), (105), and (106/107) planes, along with higher-order peaks in the 60–65° range, confirming the well-ordered layered crystalline structure of the MAX phase. After selective etching of Al, the (002) peak shifts markedly to ~5.4° in *P*-(TiVCrMo)C3, indicating significant interlayer expansion due to surface terminations and intercalation, while a weaker reflection at ~18–20° corresponds to the (004) plane. Upon oxygen functionalization (*O*-(TiVCrMo)C$_3$), the (002) peak further downshifts to ~5.17° and broadens, accompanied by suppression of high-angle reflections, indicating increased interlayer spacing and reduced crystallinity.

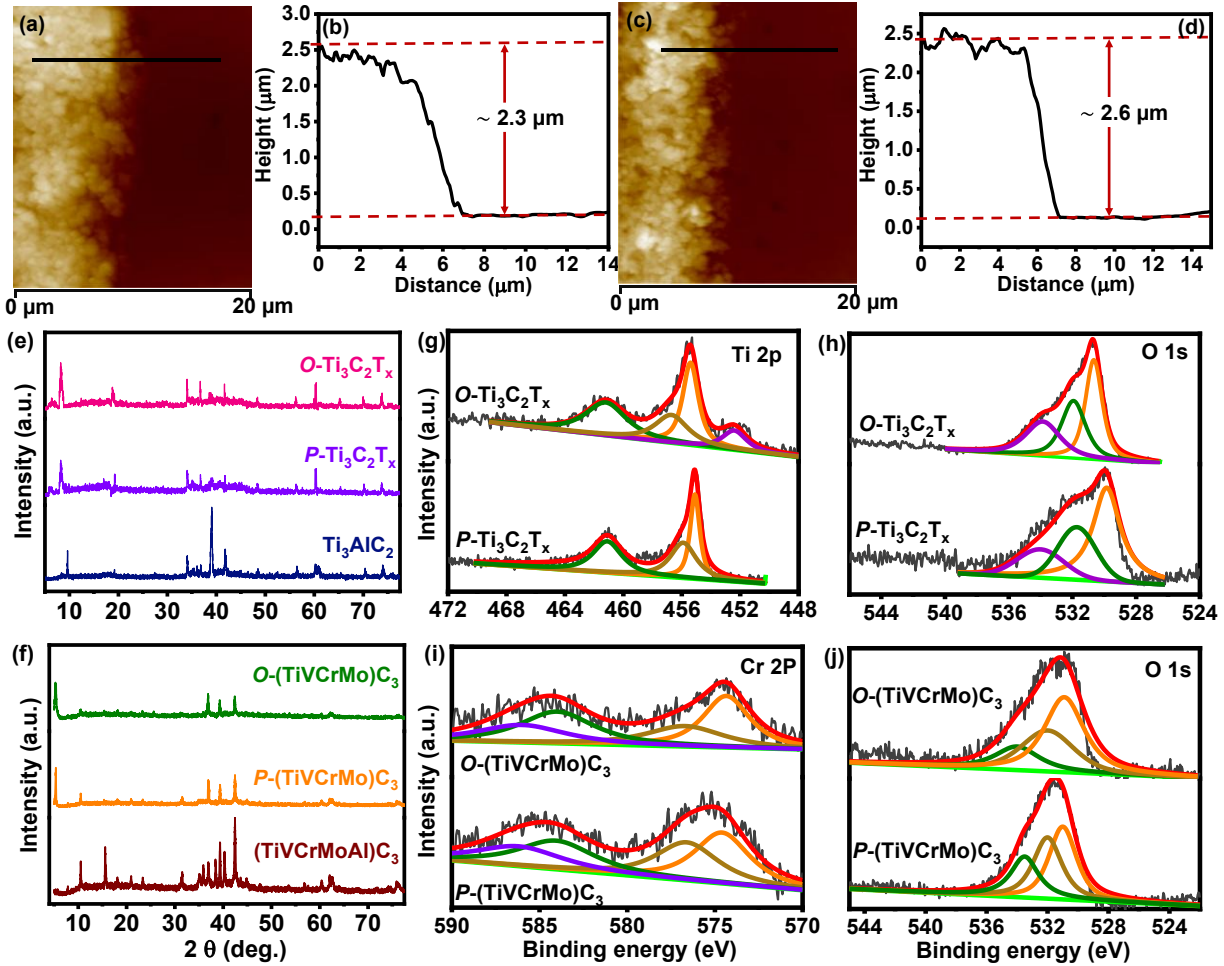

*Figure 3: AFM characterization of the MXene–epoxy interface: (a,c) surface topography image showing the step between the epoxy substrate and the deposited P-Ti$_3$C$_2$T$_x$ and P-(TiVCrMo)C$_3$ layer respectively, and (b,d) corresponding height profile. XRD plots (e) for binary MXene and (f) for medium-entropy MXene. XPS Analysis (g) high resolution spectra of Ti 2p (h) O 1s of binary MXene and (i) high resolution spectra of Cr 2p (j) high resolution spectra of O 1s of medium-entropy MXene.*

High-resolution XPS was acquired from *P*-Ti$_3$C$_2$T$_x$, and *O*-Ti$_3$C$_2$T$_X$ MXene's. Both samples in **Figure 3(g)** show a prominent carbide doublet at binding energy (BE) ~455.0/460.7 eV, confirming exposed Ti–C within the surface-near $Ti_3C_2$ layers. A pair at ~456.6/462.3 eV assigned to sub-stoichiometric $TiO_x$ /C–Ti–O($Ti^{2+}$/$Ti^{3+}$ in mixed coordination), and a high-BE pair at ~458.6/464.3 eV consistent with $Ti^{4+}$–O. $O$-$Ti_3C_2T_X$ MXene sample exhibits a marked increase in the $TiO_2$-like component, growth/broadening of the C–Ti–O contribution, and a resultant reduction of the Ti–C area. These changes clearly demonstrate surface oxidation/functionalization rather than bulk decomposition.

The O 1s spectra in **Figure 3(h)** corroborates the Ti 2p analysis. A dominant peak at ~530.0 eV ($O^{2-}$ in Ti–O) increases substantially after oxygenation, while features at ~531.2 eV assigned to –OH and C–Ti–O terminations are also intensified and slightly broadened, indicating a distribution of Ti–O bond environments (bridging O, terminal –OH). Higher-BE components at ~532.5 eV arise from molecular $H_2O$ and organic C–O species; these are more evident for the pristine surface, where the thinner oxide/termination layer exposes and traps adventitious adsorbates.

The C 1s envelopes in **Figure S4(a)** contain a low-BE C–Ti component at ~281.8 eV, a dominant C–C/C=C peak at ~284.7 eV, and oxygenated carbon at ~286.4 eV (C–O) and ~288.7 eV (C=O / O–C=O). In $O$-$Ti_3C_2T_X$ the C–Ti contribution decreases while C–O/C=O increase, reflecting conversion of some carbon to interfacial C–Ti–O and adsorption of oxygenated species as the Ti–O termination grows. The persistent 284.7 eV peak is largely adventitious/graphitic carbon, which also acts as an internal reference.

High-resolution XPS of $P$-(TiVCrMo)$C_3$, and $O$-(TiVCrMo)$C_3$ are presented in **Figure 3(i,j)**. The Cr 2p spectra is dominated by a Cr $2p_{3/2}$ feature centered at ~574.4 eV with a matched $2p_{1/2}$ peaks at ~584.2–584.5 eV. This component is attributed to Cr–C bonding in the MXene carbide lattice. Both samples also show a distinct higher-BE contribution at ~576.0 eV, assigned to $Cr^{3+}$–O. A very weak shoulder near ~579 eV is consistent with $Cr^{6+}$ species that typically form only at the surface. Relative to *P*-(TiVCrMo)C$_3$, the O-terminated exhibits a visibly larger oxide fraction and a small positive BE shift of the oxide components, indicative of a more electron-poor Cr environment after oxygenation.

The Mo region in **Figure S4(b)** resolves into two chemically distinct doublets. The lowest-BE pair at ~228.9/232.0 eV corresponds to Mo–C (carbide). A second pair at ~230.0/233.1 eV is assigned to $Mo^{4+}$–O, and a third at ~232.7/235.9 eV to $Mo^{6+}$–O. The O-terminated surface shows a pronounced increase in the oxide doublets, especially $Mo^{6+}$ at the expense of the carbide signal, whereas the pristine surface retains a larger Mo–C contribution with only modest $Mo^{4+}$/$Mo^{6+}$ content. This trend demonstrates preferential oxidation of Mo, which is the most oxophilic site.

The Ti 2p envelopes in **Figure S4(C)** contain three principal chemical states for both. The ~455.0/460.7 eV doublet is characteristic of Ti–C, confirming that the surface still exposes carbide Ti even after oxygenation. Components at ~456.7/462.4 eV (C–Ti–O) and at ~458.6/464.3 eV ($Ti^{4+}$–O) account for the oxidized Ti. In the O-sample the oxide-derived bands are clearly intensified and broadened relative to *P*-(TiVCrMo)C$_3$, consistent with a thicker, chemically heterogeneous Ti–O termination layer (mixture of O and OH). The preserved Ti–C peak evidence that oxidation remains largely a surface phenomenon and that the underlying carbide framework is intact.

The O 1s spectra in **Figure 3(j)** corroborates the metal-core analyses. All spectra show a dominant lattice-oxygen peak at ~530.0 eV assigned to M–O (M = Ti, Mo, Cr). Higher-BE components at ~531.2 eV are attributed to terminal –OH / C–M–O groups that are typical terminations for etched MXene, while ~532.5 eV features arise from adsorbed $H_2O$ and C–O/COOH species. The O-terminated displays stronger intensity in both the lattice-oxygen and –

OH/C–M–O components than the pristine sample, again reflecting more extensive oxygen termination and/or thin-oxide formation upon the oxygenation step.

## 3.2 Anticorrosion performance of the coatings

Before the EIS data was analyzed and fitted to an electric equivalent circuit (EEC), the Kramers-Kronig transforms were calculated to demonstrate the stability of the data, linearity, and causality constraints of Linear Systems Theory (LST). The good agreement between the transformed and experimental data demonstrated that the experimental impedance data was robust. The chosen EEC consists of three distinct time constants in parallel, representing electrochemical processes at different scales as shown in **Figure S6**. At the highest frequencies, the coating time constant is seen represented by ($R_{coat}$/$CPE_{coat}$), at mid frequencies, the passive film is modeled by a resistance ($R_{film}$) in parallel with a constant phase element ($CPE_{film}$), capturing both the resistive and capacitive nature of the oxide layer. At lower frequencies, the double layer capacitance ($CPE_{dl}$) and the charge transfer resistance ($R_{ct}$) describe interfacial reactions associated with metal dissolution.

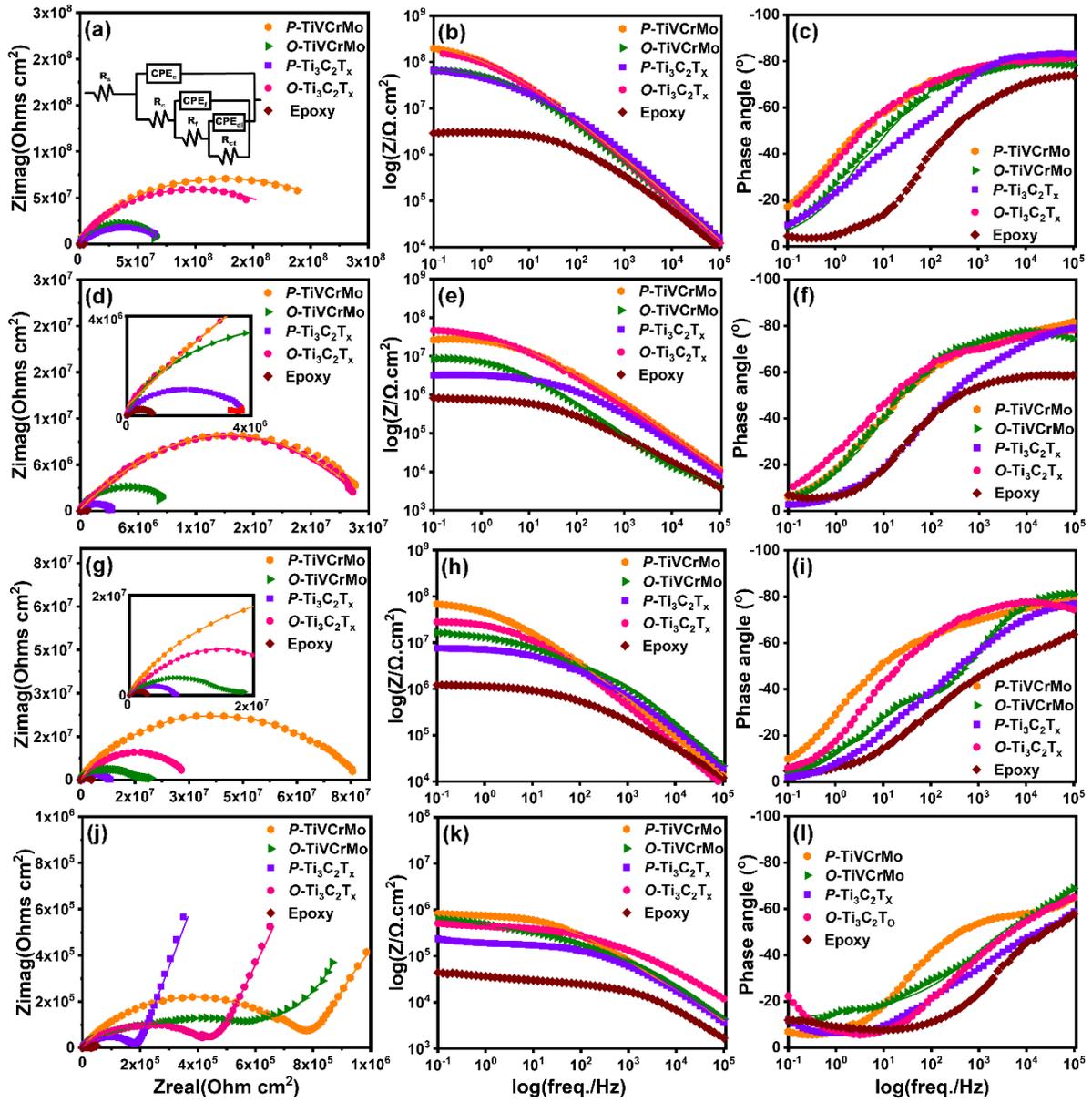

*Figure 4: Nyquist and Bode plots of various MXene–epoxy coatings P-Ti$_3$C$_2$T$_x$, and O-Ti$_3$C$_2$T$_X$, P-(TiVCrMo)C$_3$, and O-(TiVCrMo)C$_3$ recorded in 3.5 wt.% NaCl solution under different conditions: (a–c) 1 h immersion, five-layer coatings; (d–f) 1 h immersion, three-layer coatings; (g–i) 24 h immersion, five-layer coatings; (j–l) 24 h immersion, three-layer coatings.*

**Figure 4(d-f)** shows the Nyquist and bode plots of the 3 layers 1 h immersion, the *P*-(TiVCrMo)C$_3$ coating outperforms the other coatings due to the higher impedance values, followed by *O*-Ti$_3$C$_2$T$_x$, *O*-(TiVCrMo)C$_3$ and *P*-Ti$_3$C$_2$T$_x$. The shape of the Nyquist plots resemble a well-

defined semicircle, which corresponds to the low immersion time and high coating performance in the 3.5 wt.% NaCl. Similarly, the 5 layers 1 h immersion in **Figure 4(a-c)** shows the same trend; however, the overall impedance values have increased by one order of magnitude. This is attributed to the higher thickness due to the low immersion time; thus, no water uptake is seen in the coating [21]. The corrosion performance of MXene-based coatings is strongly governed by their surface chemistry and lattice configuration, which together dictate ion adsorption, charge transfer, and structural stability [22]. In binary $Ti_3C_2T_x$ MXene, mild oxygenation substantially enhances anticorrosion behavior by replacing inert –F terminations with polar –O and –OH groups that promote chemical activity and interfacial bonding with the epoxy matrix. The oxygenated surface provides abundant adsorption sites for corrosive $Cl^-$ and reactive $Fe^{2+}/Fe^{3+}$ species, effectively immobilizing them within the coating and suppressing localized electrochemical reactions. Simultaneously, improved dispersion and interlayer compaction generate a more tortuous diffusion path for electrolyte ingress, while the inherent metallic conductivity of MXene facilitates charge redistribution between anodic and cathodic microcells, reducing localized potential gradients. These combined effects transform the $O$-$Ti_3C_2T_x$ from a passive filler into an active barrier-inhibitor hybrid, yielding coating resistances on the order of $10^7$ $\Omega cm^2$ an order of magnitude higher than its pristine counterpart.

In contrast, the corrosion behavior of the medium-entropy $(TiVCrMo)C_3$ MXene follows a different mechanism due to its configurationally stabilized lattice and multi-metal redox functionality. The $P$-$(TiVCrMo)C_3$ exhibits superior protection through dense lamellar stacking, high electronic conductivity, and in-situ passivation driven by Cr- and Mo-rich oxide formation under exposure, resulting in self-healing behavior akin to stainless steels. Oxygenation, however, partially disrupts this equilibrium by generating amorphous surface oxides that reduce conductivity, hinder redox adaptability, and introduce interfacial defects, thereby diminishing the overall impedance response. Consequently, the corrosion resistance follows the order *P*-(TiVCrMo)C$_3$ > *O*-Ti$_3$C$_2$T$_X$ > *O*-(TiVCrMo)C$_3$ > *P*-Ti$_3$C$_2$T$_X$ underscoring that while surface oxygenation enhances protection in binary MXene's via adsorption and sealing, it can impair the entropy-driven stability and dynamic passivation intrinsic to multicomponent MXene's.

Once the immersion time increases to 24 hours as depicted in **Figure 4(g-l)**, the coating has been able to gain some water uptake, filling the pores of the MXene-epoxy layers and making them more conductive. This is reflected in the decrease of two orders of magnitude in the impedance from $10^7$ to $10^5$ $\Omega$cm$^2$. Nevertheless, even if the substrate is carbon steel, the impedance is high proving that the coating is still performing, having a high impedance value.

Regardless of the higher immersion time, the 5 layers were able to remain close to the original values, at least for the *P*-(TiVCrMo)C$_3$, proving again the outperforming capabilities against corrosion in comparison to the other coatings. Since the 5 layers had an increase in thickness of around 30-50 μm, and both *P*-(TiVCrMo)C$_3$ and *O*-Ti$_3$C$_2$T$_X$ were the best coating from the 4 selected for this study, two samples were done a scratch and immersed for 24 hours to do impedance analysis and see how the coating layers would react **Figure S5(a-c)**. The reduction of impedance to approximately $10^5$ $\Omega$cm$^2$ was observed after scratch can be attributed to the partial mechanical disruption of the layered epoxy-MXene architecture, which exposes localized regions of the underlying steel substrate and creates direct electrolyte pathways through the coating. In the intact state, the five-layer structure provides both a tortuous diffusion path and strong interfacial sealing, but the scratch introduces microcracks and delamination that accelerate electrolyte penetration and charge transfer across the coating metal interface as can be seen from the SEM-EDS images taken after the scratch test depicted in **Figure 5**. Although the overall impedance decreases by nearly three orders of magnitude, the residual resistance remaining in the $10^5$ $\Omega cm^2$ range indicates that the MXene on the walls of the scratch can still react with the electrolyte and hinder the chloride ingress. Moreover, MXene near the substrate promotes the formation of a protective oxide layer, composed primarily of $Fe_2O_3$ and $Fe_3O_4$ thus, to some extent shields the substrate from corrosion [22]. As seen with the work of Zhao et al.[23], through scratch testing, MXene in close proximity to the exposed metal promotes the formation of a protective oxide film, which serves as the primary mechanism for corrosion inhibition. While the intact coating had a tortuous path for the chlorides to reach the substrate, with the scratch, that mechanism is not as effective due to the free path of the chlorides in the solution to reach the substrate.

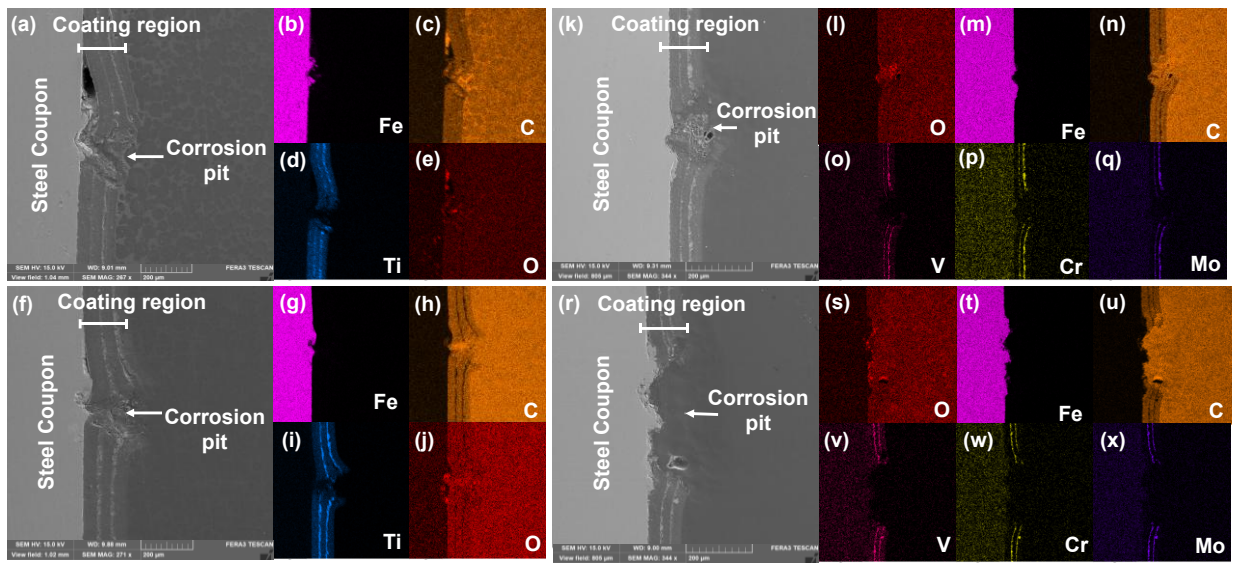

*Figure 5: SEM micrographs and corresponding EDS analyses of scratched multilayer coatings after EIS testing: (a) cross-sectional view showing the five-layer architecture of P-$Ti_3C_2T_x$ with (b–e) corresponding EDS elemental maps; (f) cross-sectional view of the O-$Ti_3C_2T_x$ layer with (g–j) elemental distributions; (k) cross-sectional SEM image of the P-$(TiVCrMo)C_3$ coating with (l–q) corresponding EDS maps; and (r) cross-sectional view of the O-$(TiVCrMo)C_3$ layer with (s–x) EDS elemental mapping.*

All the fitted components of the EEC can be found in **Figure S7**. Given the large amount of resistance and CPE, the key elements will be discussed in depth, while the others will only be roughly compared. As expected from a thick coating, the $CPE_{coat}$ was low in the $10^{-8}$ for 3 layers, further decreasing to $10^{-10}$ for 5 layers, Similarly, the corresponding exponent (n) approaches unity, indicating near-ideal capacitive behavior for thicker coatings, even after 24 h of immersion. Following scratching, n decreases to 0.78–0.87, still suggesting a relatively low defect density. This behavior supports the hypothesis that, upon water ingress, the multilayer structure interacts with the electrolyte to hinder the penetration of aggressive ions. The $R_{coat}$ exemplifies the effect of the thickness and how each coating varies in mechanism. The *P*-(TiVCrMo)C$_3$ and *O*-Ti$_3$C$_2$T$_X$ increase about 2 orders of magnitude, which, by only adding one layer of the respective MXene, shows how crucial its addition is to the enhancement of the corrosion performance. First due to the higher thickness, thus longer distance for the water and ions to diffuse, and second, more concentration of the MXene in the coating, thus more possible interactions to react with the ions once inside to trap them. The 24-hour immersion did not change severely the corrosion performance of the coating, coinciding with the slight change in impedance from the Nyquist. It was not until the scratch was made that the $R_{coat}$ of 5 layers after 24 hours of immersion reduced one order of magnitude. Which still shows higher value than the 3-layer proving that the MXene can react/heal at the scratch walls. The Rct follows the same trend seen for the Rcoat, as well as the overall trend where from more to less protective against corrosion *P*-(TiVCrMo)C$_3$ > *O*-Ti$_3$C$_2$T$_X$ > *O*-(TiVCrMo)C$_3$ > *P*-Ti$_3$C$_2$T$_x$. Since the values of the Rct were above the 100 k $\Omega$cm$^2$ for all cases and even in the 10 M $\Omega$cm$^2$ for the *P*-(TiVCrMo)C$_3$ after the 24h immersion with the scratch, this shows that the coating can provide extremely good corrosion protection, and that the bare substrate does not see the effect of the chlorides.

### 3.3 ELF analysis

ELF analysis in **Figure 6(a-d)** and **Figure S8** reveals that the carbon atoms in pristine and oxygen-functionalized MXene exhibit pronounced electron localization, as indicated by the red regions surrounding C atoms. This behavior is the same in high-entropy MXene, suggesting a local charge redistribution induced by multi-metal environments and strong interaction with oxygen species.

In contrast, after water adsorption, the ELF intensity around carbon atoms decreases to orange levels, indicating reduced electron localization. Water adsorption mainly modifies the local electrostatic environment and bonding configuration at the surface. The interaction between water molecules and surface metal or oxygen sites screens the surface polarization and weakens the directional character of C–M bonding, leading to more delocalized electronic states around carbon atoms, as reflected by the reduced ELF intensity.

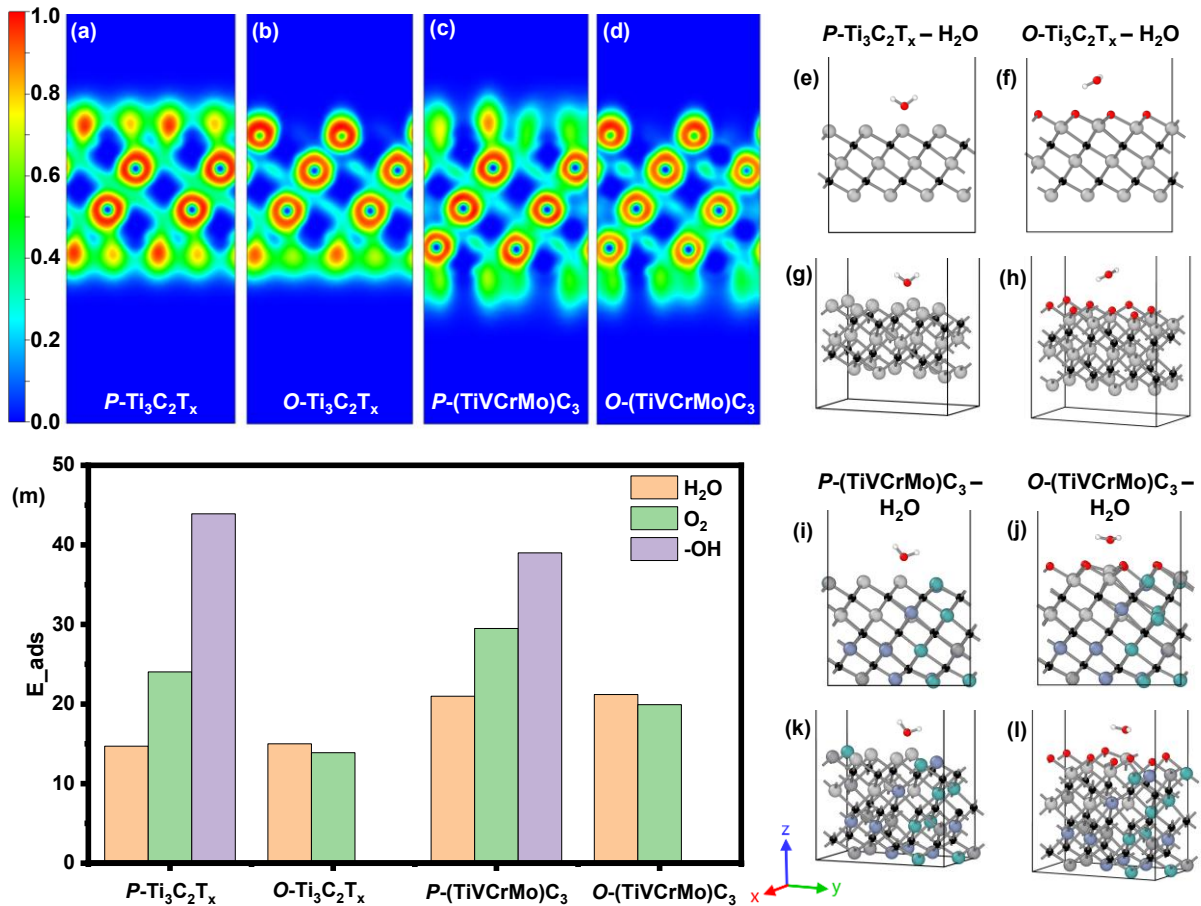

*Figure 6: (a-d) Electron localization function (ELF), (e-l) optimized adsorption configurations, (m) adsorption energies of oxygen-containing species ($H_2O$, $O_2$, and OH).*

In contrast to oxygen functional groups that induce strong charge polarization and enhance electron localization within the C–M (M= Ti, Cr, Mo, and V) bonding network, water adsorption primarily alters the surface electrostatic field and bonding environment through weak interactions, resulting in a more delocalized electronic distribution around carbon.

The adsorption of water molecules does not directly participate in surface bond reconstruction but instead perturbs the local electrostatic potential near the surface. This perturbation weakens the polarization of metal–carbon bonds induced by oxygen species, causing the bonding electrons to redistribute into more delocalized states. As a result, the electron localization around carbon atoms is suppressed, which may influence the initiation and kinetics of subsequent oxidation reactions.

For the binary $Ti_3C_2$ system, (**Figure 6(m), Figure S9**) the pristine surface exhibits very strong $O_2$ adsorption, indicating high surface reactivity and susceptibility to oxygen-assisted interfacial reactions. Oxygen functionalization partially saturates the exposed Ti sites, weakening further $O_2$ adsorption and stabilizing the surface, which is consistent with the improved corrosion resistance observed for *O*-$Ti_3C_2T_x$ compared with *P*-$Ti_3C_2T_x$. In contrast, the *P*-$(TiVCrMo)C_3$ MXene exhibits inherently stronger adsorption toward $O_2$, $H_2O$, and −OH due to the presence of multiple transition metals that create heterogeneous electronic environments and abundant active sites. This strong adsorption facilitates interaction with oxygen-containing species and promotes the formation of stable protective oxides such as $Cr_2O_3$ and $MoO_3$, enabling dynamic passivation of the surface. While in case of *O*-$(TiVCrMo)C_3$, partial saturation of the active metal sites reduces adsorption strength and limits the redox activity necessary for passivation, leading to lower corrosion resistance compared with *P*-$(TiVCrMo)C_3$. Overall, the DFT results therefore support the experimental performance trend.

## 4   Conclusion

This work demonstrates that corrosion protection in MXene-based multilayer coatings is dictated by the coupled effects of surface termination, electronic structure, and compositional entropy. Oxygen functionalization improves the performance of binary $Ti_3C_2T_x$ by stabilizing reactive Ti sites, strengthening interfacial sealing, and enhancing ion-trapping capability, whereas the same treatment weakens medium-entropy $(TiVCrMo)C_3$ by partially saturating multi-metal active sites and reducing the redox adaptability required for dynamic passivation. Density functional theory calculations support this distinction, showing that oxygenated $Ti_3C_2$ surfaces are less prone to further oxygen-driven reactivity, while $P$-$(TiVCrMo)C_3$ retains strong adsorption toward $O_2$, $H_2O$, and $-OH$ due to its heterogeneous multi-metal electronic environment, favoring the formation of protective Cr and Mo rich oxides. These results explain the observed performance hierarchy and establish that corrosion behavior in MXene coatings is fundamentally composition-dependent, providing mechanistic design principles for next-generation anticorrosion nanocomposites.

**Supporting Information :** Link to be provided by Editor.

**Author Information**

**Corresponding Authors**

Chenglin Wu − Department of Civil and Environmental Engineering, Texas A&M University, College Station, TX, 77843, USA

**Authors**

Aqsa Fayyaz − Department of Civil and Environmental Engineering, Texas A&M University, College Station, TX, 77843, USA

Ulises Martin Diaz − Department of Materials Science and Engineering, Texas A&M University, College Station, TX, 77843, USA

Jianyu Dai − Department of Civil and Environmental Engineering, Texas A&M University, College Station, TX, 77843, USA

Homero Castaneda − Department of Materials Science and Engineering, Texas A&M University, College Station, TX, 77843, USA

**Author Contributions**

Aqsa Fayyaz synthesized and delaminated four MXene $P$-(TiVCrMo)C$_3$, $O$-(TiVCrMo)C$_3$, $P$-Ti$_3$C$_2$T$_x$, $O$-Ti$_3$C$_2$T$_x$, fabricated multilayer coatings on carbon steel, carried out electrochemical testing to assess anticorrosion performance, performed material characterizations, curated and analyzed the resulting data, and prepared the manuscript. Ulises Martin Diaz assisted with the experimental work, evaluated the EIS results, and contributed to drafting the corresponding section of the manuscript. Jianyu carried out the DFT calculations and performed the ELF analysis. Chenglin Wu provided supervision, conceptualized the project, conducted theoretical modeling, reviewed the manuscript, and secured funding. Homero Castaneda provided supervision on the electrochemical experiment, analysis, and reviewed the manuscript.

**Acknowledgments**

The authors gratefully acknowledge support from the National Science Foundation CMMI #2414716. They also appreciate the use of the Materials Characterization Facility at Texas A&M University (RRID: SCR_022202).